\documentclass[preprint,amsmath,amssymb,prb,floatfix]{revtex4-2}
\usepackage{graphicx}
\usepackage{dcolumn}
\usepackage{bm}
\usepackage{csquotes}
\usepackage{graphics}
\usepackage{physics}
\usepackage{float}
\usepackage{nicefrac}
\usepackage{xspace}
\usepackage{bm}
\usepackage{titlesec}
\usepackage{akkmathset}
\usepackage{soul}
\usepackage{float}
\usepackage{tabto}
\usepackage{array}
\usepackage{makecell}
\usepackage{nicefrac}
\usepackage[utf8]{inputenc}
\usepackage[T1]{fontenc}
\usepackage{mathptmx}
\usepackage{amsmath,amssymb,mathtools}
\usepackage{multirow}
\newcommand{\ceq}[1] {(\ref{#1})}

\newcommand{\nbse}{NbSe$_2$\xspace}
\newcommand{\mos}{MoS$_2$\xspace}

\newcommand{\nbs}{NbS$_2$\xspace}
\newcommand{\pdte}{PdTe$_2$\xspace}
\newcommand{\ssb}   {{\bf S}}

\newcommand{\bR}     {{\bf R}}
\newcommand{\kk}     {{\bf k}}

\newcommand{\beginsupportinginfo}{%
        \setcounter{figure}{0}
        \renewcommand{\thefigure}{S\arabic{figure}}%
        \setcounter{equation}{0}
        \renewcommand{\theequation}{S\arabic{equation}}%
     }

\begin{document}

\title{Kondo Effect in Defect-bound Quantum Dots Coupled to \nbse}
\author{T. R. Devidas}
\author{Tom Dvir}
\affiliation{The Racah Institute of Physics, The Center for Nanoscience and Nanotechnology,  The Hebrew University, Jerusalem 91904, Israel}

\author{Enrico Rossi}
\affiliation{Department of Physics, William \& Mary, Williamsburg, VA 23187, USA}

\author{Hadar Steinberg}
\email{hadar@phys.huji.ac.il}
\affiliation{The Racah Institute of Physics, The Center for Nanoscience and Nanotechnology,  The Hebrew University, Jerusalem 91904, Israel}

\begin{abstract}
\begin{quotation}

We report the fabrication of a van der Waals tunneling device hosting a defect-bound quantum dot coupled to NbSe$_2$.
We find that upon application of magnetic field, the device exhibits a zero-bias conductance peak. The peak, which splits at higher fields, is associated with a Kondo effect.
At the same time, the junction retains conventional quasiparticle tunneling features at finite bias.
Such coexistence of a superconducting gap and a Kondo effect are unusual, and are explained by noting the two-gap nature of the 
superconducting state of \nbse, where a magnetic field suppresses the low energy gap associated with the Se band.
Our data shows that van der Waals architectures, and defect-bound dots in them, can serve as a novel and effective platform for investigating  the interplay of Kondo screening and superconducting pairing in  unconventional superconductors.    
  
\end{quotation}
\end{abstract}

\maketitle

\section{Introduction}
The Kondo effect~\cite{Kondo1964}~ is responsible for the low temperature resistivity upturn in metals with dilute magnetic impurities~\cite{DeHaas1934}~, and is a paradigmatic problem in Condensed Matter Physics. 
In the Kondo effect, the coupling of a single spin to a metallic environment causes the formation of a magnetic screening cloud. 
It embodies the complexity arising when a single particle interacts with a many-body environment, and was studied extensively using electronic transport through GaAs quantum dots (QDs)~\cite{Goldhaber-Gordon1998}~, and later carbon nanotubes~\cite{Nygard2000}~ and single molecules~\cite{Park2002}~. In these systems, the effect manifests itself as an enhancement of the electronic conductance due to correlations between the source and drain mediated by an unpaired spin localized within the QD.

A natural extension of the Kondo problem involves coupling a QD to one or two superconducting leads. 
The first experimental study where a QD in the Kondo regime was coupled to superconducting (S) source and drain electrodes was reported by Buitelaar et al. ~\cite{Buitelaar2002}~. In this `S-QD-S' geometry, the electron number in the impurity level was tuned to an odd value via electronic gating. The SU(2) Kondo signature – zero-energy state - was observed whenever the Kondo temperature $T_K$ became the dominant energy scale in the system. 
Such an observation falls in line with the theory of many-body singlet state for the Kondo ground state, which can create a Kondo signature whenever the binding energy of the Kondo singlet exceeds the superconducting energy gap $\Delta$~\cite{Clerk2000}~. 

Later studies have shown that even when the zero-bias Kondo feature is  suppressed, an applied source-drain bias $V_{SD} = \Delta$ reveals the formation of a Kondo resonance within the QD - the strength of which appears to follow a scaling law defined by the competition between $\Delta$ and $k_{B}T_{K}$~\cite{Buizert2007,Clerk2000}~, where $k_B$ is the Boltzmann constant. 
This competition underlies a rich set of physical phenomena involving the Andreev reflection at the QD-S interface and the Kondo singlet formation at the QD-N (normal) interface ~\cite{Clerk2000,Yeyati2003,Sand-Jespersen2007}~. Theoretical studies have suggested that adopting a hybrid N-QD-S geometry would allow a more rigorous approach in studying these dynamics ~\cite{Clerk2000,Cuevas2001,Li2016,Zitko2015}~.

Experimental realizations of N-QD-S devices typically rely on clean semiconductor nanowires (GaAs, InAs)~\cite{Jespersen2006}~, coupled to superconductor materials that could be evaporated as electrodes (Pb, Nb, Al, MoRe). 
The advent of van der Waals (vdW) materials now enables the study of a whole library of materials with diverse ground states ~\cite{Ajayan2016}~. Specifically, the Transition Metal Dichalcogenides (TMD) family contains candidate 2D superconductors (\nbse, \nbs, \pdte) that exhibit superconductivity from the bulk regime to the monolayer limit ~\cite{Ugeda2016}~. 
Insulators such as \mos, \text{WSe$_2$}, and hexagonal Boron Nitride (hBN) have been successfully implemented as barriers in tunnel junctions ~\cite{Dvir2018a, Dvir2018b, khestanova2018}~ and have been shown to host atomic defects ~\cite{Zhou2013, Dvir2019, Devidas2021}~ which could be used as QDs ~\cite{Chandni2015, Greenaway2018, Keren2020}~. 

In this work we demonstrate that a Kondo effect can also be realized by coupling a QD to a layered superconductor. To place a QD in close proximity to a layered superconductor, we make use of naturally-occurring defects in a TMD semiconductor tunnel barrier placed on top of the superconductor \nbse. As we have shown in previous studies, such defect-bound dots may couple strongly to the SC, giving rise to Andreev bound state (ABS) sub-gap conductance features~\cite{Dvir2019}~. Conversely, when the QD is weakly coupled, it can serve as a sensitive spectral~\cite{Devidas2021}~ or compressibility~\cite{Keren2020}~ probe.
In the present case, we find a tell-tale zero-bias conductance peak which sets in at finite in-plane and out-of-plane magnetic field. The peak splits at higher magnetic fields, consistent with a Kondo feature.
In \nbse, tunneling measurements resolve two gaps~\cite{Noat2015,Dvir2018a}~. Here we find that the Kondo feature is correlated with a suppression in the spectral signature of the lower of these two energy gaps, associated with the Se band.
The correlation of the peak value with the suppression of the second gap tunneling feature suggests that the Kondo effect is mediated by carriers belonging to the Se-derived band, which turns normal at a low magnetic field, while the larger Nb-derived gap remains stable. Our results suggest that defect-bound QDs can be used to study the Kondo effect in unconventional, van der Waals, layered superconductors.

\section{Experimental Details}
\begin{figure}
    \centering
    \includegraphics[width = 400 pt]{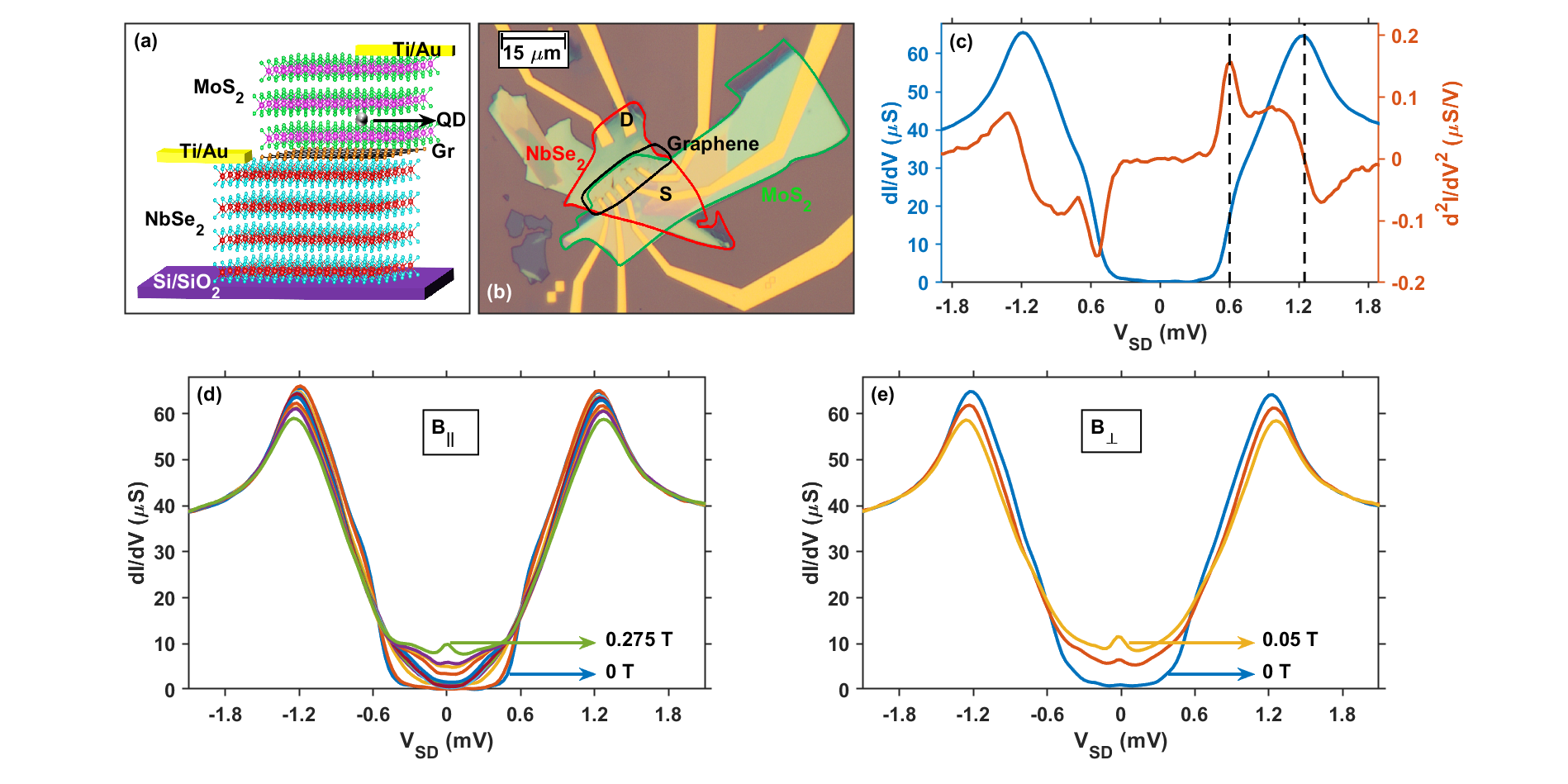}
    \caption{(a) Schematic of the device measured in the current work; (b) Optical image of the measured device. \nbse, graphene and \mos flakes are outlined and labelled in different colours for easier identification. S and D indicate the source and drain electrodes respectively. (c) Left $y$-axis - Differential conductance ($dI/dV$) spectrum of the \nbse-graphene-\mos tunnel junction depicted in (b), Right $y$-axis - Derivative of the differential conductance ($d^2I/dV^2$) spectrum measured at 28 mK. The black dashed lines track the two \nbse gaps present in the spectrum at 0.6 mV and 1.25 mV; (d) and (e) show the evolution of the tunneling spectrum as a function of parallel ($B_{||}$) and perpendicular magnetic field ($B_{\perp}$) respectively, for low magnetic field values. }
   \label{fig:fig1}
\end{figure}

The schematic of the \mos-graphene-\nbse vdW heterostructure is shown in Figure \ref{fig:fig1}(a) and the optical image of the measured device reported in this work is shown in Figure \ref{fig:fig1}(b). 
\mos, graphene and \nbse are exfoliated independently on Si/SiO$_2$ (285 nm oxide) substrates. The desired flakes of 3-4 layers \mos (tunnel barrier), monolayer graphene and bulk \nbse are chosen by optical contrast. The \mos barrier flake is first picked up using the polycarbonate (PC) technique. Graphene is subsequently picked up using the van der Waals interaction between \mos and graphene. The picked up heterostructure is then transferred on to a 25 nm thick bulk \nbse flake. The process is carried out in a glovebox under an Argon atmosphere. Standard e-beam lithographic techniques are used to pattern tunnel electrodes and ohmic contacts on \mos and \nbse respectively. Ti/Au electrodes are evaporated using an e-beam evaporator with an additional Argon ion milling step prior to the ohmic contact deposition step, so as to obtain better contact on the \nbse flake. No contacts are made to the graphene flake.

\section{Results}

The tunneling differential conductance ($dI/dV$) of the \mos-graphene-\nbse\ stack at 28 mK and B = 0 T, is shown on the left $y$-axis of Figure 1(c). The spectrum exhibits well-defined quasiparticle peaks and a hard gap - manifest in a ratio of over 100 between the conductance outside the gap ($G_N$) and zero-bias conductance ($G_0$). Such a hard gap, seen in our earlier studies on \nbse tunneling devices~\cite{Dvir2018a,Dvir2018b}~, attests to the quality of the tunnel junction, which suppresses two-particle tunneling contributions. 
The $2^{nd}$ derivative of the tunneling current ($d^2I/dV^2$), plotted in the right $y$-axis of Figure 1(c) shows a clear separation between two distinct features marked by black dashed lines. These two features correspond to the higher energy gap, $\Delta_1 = 1.25$ mV, and the lower energy gap $\Delta_2 = 0.6$ mV. These were identified in the past as related to the two superconducting bands in \nbse – the 2D-like niobium $d$-orbitals bands around the $\Gamma, K$ points, associated with $\Delta_1$, and selenium-derived $p$-orbitals which form a small 3D Fermi surface around the $\Gamma$-point, associated with $\Delta_2$ ~\cite{Dvir2018a,Dvir2018b,Kiss2007,Noat2015}~. 

Interestingly, the tunneling spectrum does not reveal any signature which can be associated with the graphene layer.
This can be understood considering that only for a small range of twist angles graphene's Fermi pockets 
can have some overlap with the Fermi pockets of a TMD like \nbse~\cite{Gani_NbSe2_Gr_2019}~.
As a result, the graphene spectrum may be decoupled from the \nbse. 

Figure 1(d) shows the evolution of the tunneling spectra as a function of in-plane magnetic field ($B_\parallel$) applied parallel to the sample plane. The observed spectra evolve in a way which is not characteristic of \nbse devices measured in the past ~\cite{Dvir2018a}~: First, we find that the superconducting gap is filled even by the application of a very low field. Second, a zero-bias conductance peak (ZBCP) develops at $B_\parallel = 0.275$ T. In what follows, we discuss possible mechanisms for these features.

Zero-bias spectral features are rather common in proximity superconducting devices, and may have a number of possible origins.
They could be associated with Andreev bound states (ABS) residing on the surface of a $d$-wave superconductor ~\cite{Millo2018}~, or with constructive superposition of bound states formed by reflectionless tunneling at diffusive N-S junctions ~\cite{VanWees1992,Marmorkos1993}~.

Zero-bias states may appear when a QD, proximity-coupled to  a superconductor, forms a local Andreev bound state which undergoes a singlet-doublet transition~\cite{Lee2014,Scherubl2020}~.  
Conversely, when the QD is weakly coupled to both SC ($\Gamma_{SC}$) and normal ($\Gamma_N$) leads, i.e.
$\Gamma_{SC,N} << \Delta$, 
QD-SC transport is dominated by single electron resonant tunneling. 
In this regime, the spectrum exhibits sharp conductance peaks, but these can appear both above the gap and below it,
when the density of states (DOS) is not zero.
In a recent publication~\cite{Devidas2021}~ we have reported the use of such a weakly-coupled QD as a sensitive spectrometer.

The zero-bias conductance peak also appears as a response to out-of-plane magnetic field $B_{\perp}$ as seen in Figure 1(e). To further study the nature of this peak, we track it's evolution as magnetic fields are applied in directions parallel and perpendicular to the sample plane.
Figure 2(a) represents a color scale map of the tunneling spectra vs. $B_\parallel$ extending to 9 T. Following the evolution of the zero-energy state that switches on at 0.275 T (Figure 1(d)), the peak persists at the same energy until a magnetic field induced splitting of the state into two distinct features is observed at 1 T. 
The splitting is symmetrical in $V_{SD}$ with respect to zero. 

Although such linear dispersions are regularly observed in ABS, here we argue that these features are unlikely to be related with ABS, on two grounds. First, ABS features in \nbse usually appear at finite bias, and may converge to zero at finite field. They are observed starting at zero magnetic field and do not require a finite field to be visible. Second, ABS features in \nbse tunnel devices are observed only in very thin flakes. In any flake thicker than a few layers, the in-plane magnetic field introduces a sub-gap tunneling signal which obscures them ~\cite{Dvir2019}~.

We are thus led to propose that the observed feature - the zero-bias conductance peak at a finite field, is associated with a Kondo origin.

The split states are separated by 2$E_Z$ ~\cite{Goldhaber-Gordon1998,Pustilnik2004}~, $E_Z=\pm g\mu_B B$ being the Zeeman energy, with a Land\'e $g$-factor (1.67$\pm$0.04). 
A similar trend is observed in the evolution of the spectrum in $B_\perp$ with a field induced splitting appearing at 0.6 T, shown in Figure 2(b). The value of Land\'e $g$-factor obtained from out-of-plane data is 1.94$\pm$0.04. We note that similar values have been seen in the past for atomic defect QDs~\cite{Dvir2019,Devidas2021}~.

\begin{figure}
    \centering
    \includegraphics[width = 420 pt]{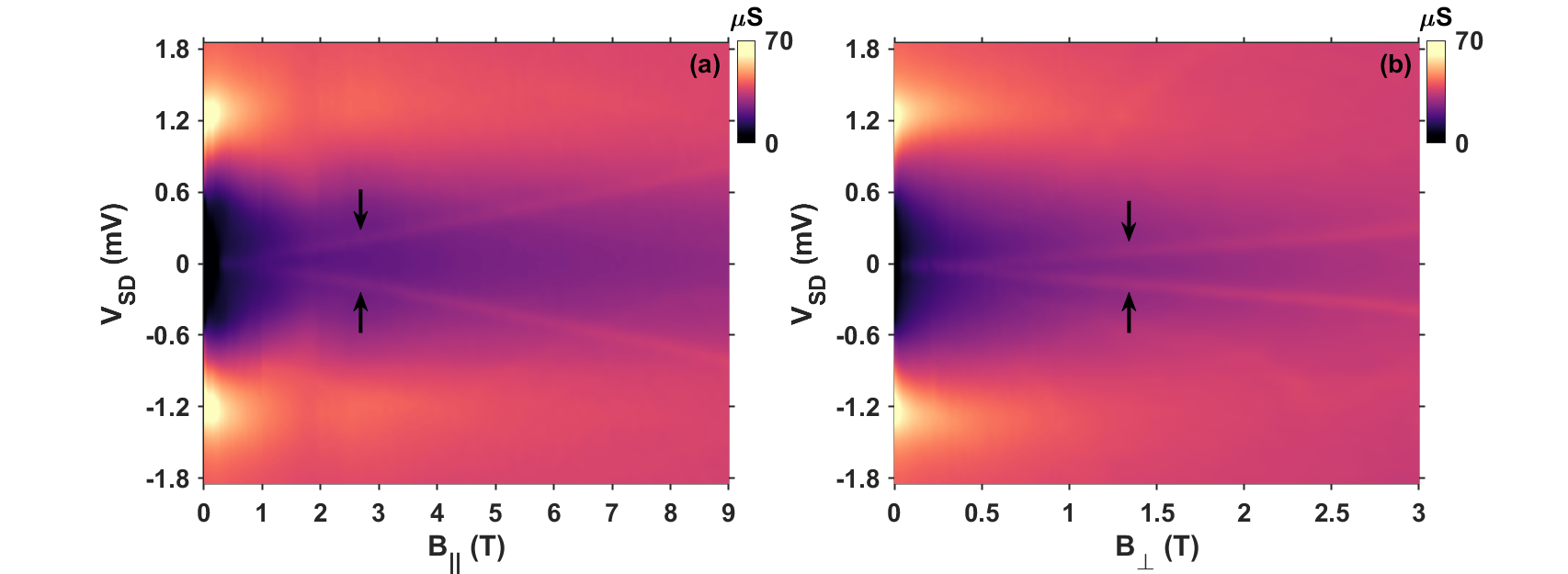}
    \caption{Color map of the $dI/dV$ spectrum in (a) in-plane magnetic field ($B_{||}$) and (b) out-of-plane magnetic field ($B_\perp$). The black arrows in (a) and (b) indicate the Zeeman split features.}
    \label{fig:fig2}
\end{figure}

The Kondo effect sets in below a Kondo temperature $T_K$ which can be obtained through temperature dependence or through the profile of the zero-bias conductance peak.
Temperature dependence measurements of the zero-bias conductance peak ($B_\perp$ = 25 mT) are carried out from 28 mK to 419 mK, beyond which the peak is no longer detectable above the background. A selection of these plots is shown in Figure 3(a). The maximal zero-bias conductance at 28 mK is 11.1 $\mu S$ = 0.14 $e^2/h$. Figure 3(b) shows a semi-log plot of zero-bias conductance peak heights as a function of temperature, where peak height values are obtained after suitable background subtraction. 
The data is fit to an empirical equation \eqref{eq1} derived from the numerical renormalization group (NRG) theory for the Kondo ground state~\cite{Buitelaar2002,Goldhaber-Gordon1998}~.

\begin{equation}
\label{eq1}
\centering
    G(T)=\frac{G_{max}}{\bigg[1+\bigg (2^{\nicefrac{1}{s}}-1\bigg)\bigg(\frac{T}{T_K} \bigg)^2 \bigg]^s}
\end{equation}

Where $G_{max}$ is the maximum conductance at the lowest temperature measured, $T_K$ is the Kondo temperature and $s$ is a dimensionless value related to the spin-state of the electron in the QD. The best fit to the data (solid red line) yields $G_{max}$ = 0.0695$\pm$0.003 $e^2/h$; $T_K$ = 153$\pm$6 mK (13.2 $\mu$eV) and $s$ = 1.4$\pm$0.3. 

Clearly, the exponent $s$ deviates from the value $s$ = 0.22, expected for a spin 1/2 system at the Kondo regime~\cite{Goldhaber-Gordon1998,Buitelaar2002}~. Values removed from 0.22 could indicate that the QD is in the mixed-valence regime, which is seen when it's energy is close to resonance, or, alternatively, that the spin is not 1/2.

\begin{figure}[ht]
    \centering
    \includegraphics[width = 400 pt]{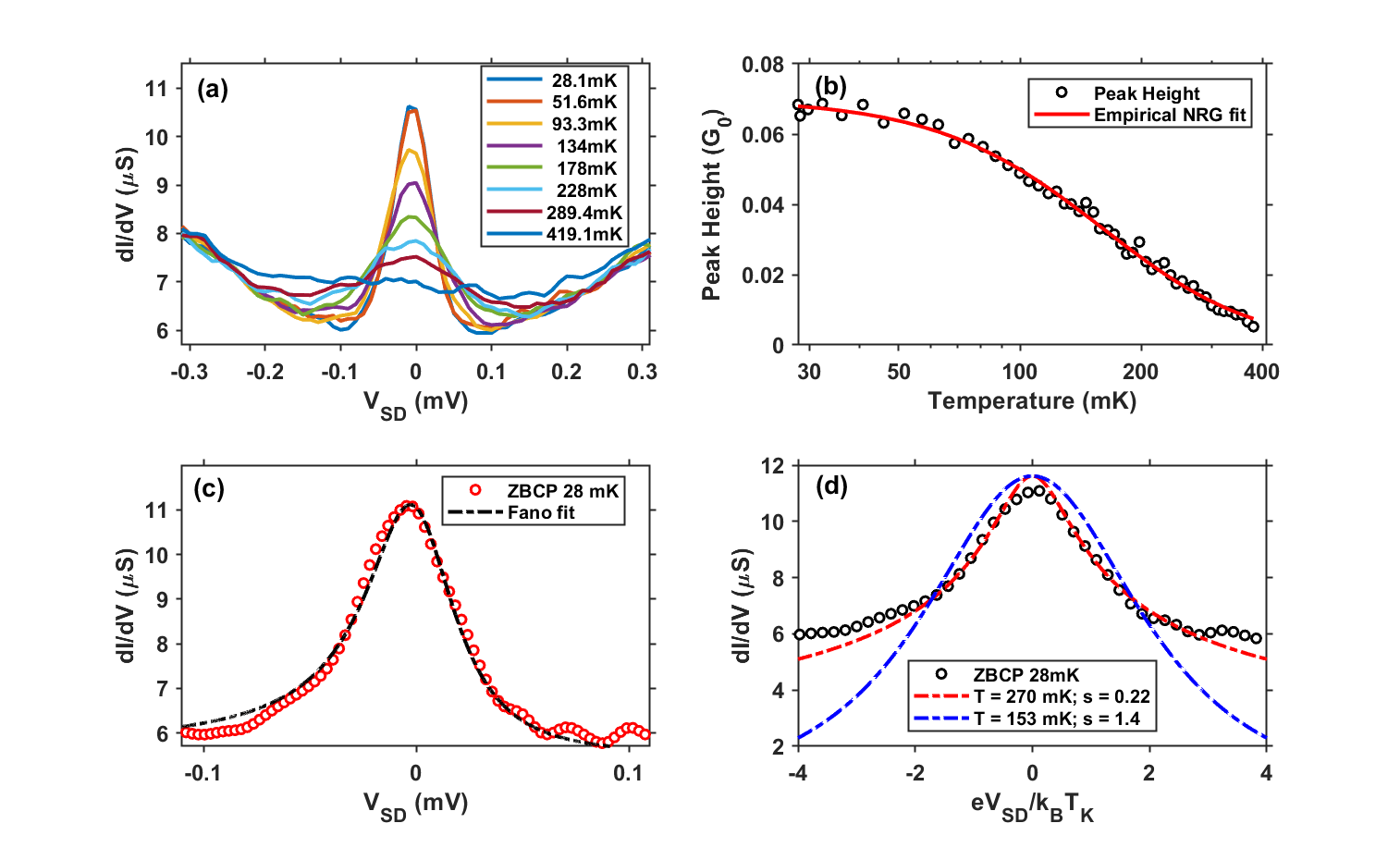}
    \caption{Evolution of the sub-gap conductance ($dI/dV$) spectra as a function of temperature in the QD-\nbse junction (a) Scans at select temperatures; (b) Kondo peak heights as a function of temperature, after suitable background subtraction, fitted (red solid line) to the NRG derived empirical equation  ~\cite{Buitelaar2002,Goldhaber-Gordon1998}~; (c) The zero-bias conductance peak at 28 mK, fit to the Fano function ~\cite{Fano1961}~ to extract the value of HWHM which defines $T_K$ ~\cite{Deacon2010b}~; (d) The zero-bias  conductance peak at 28 mK, fit to the equation derived by Kretinin et al. ~\cite{Kretinin2012}~ to different values of $T_K$ and $s$.} 
    \label{fig:fig3}
\end{figure}

It is also possible to evaluate $T_K$ from the conductance feature line-shape.
In a N-QD-S system, the half-width at half maximum (HWHM) of the peak at the lowest temperature is directly related to $T_K$ ~\cite{Deacon2010b}~. Since the Kondo state involves an interaction between a discrete state (QD) and continuum (N lead), we use the Fano function ~\cite{Fano1961}~ to extract the HWHM (Figure 3(c)). The function is defined as

\begin{equation}
\label{eq_Fano}
\centering
    G(V_{SD})= A \frac{(\varepsilon+q)^2}{(1+\varepsilon^2 )}+B; \varepsilon= \frac{(V_{SD}-\varepsilon_0 )}{\Gamma_d} 
\end{equation}

Where $\varepsilon_0$ is the resonance energy (0 eV in the present data), A and B are constants, $\Gamma_d$ is the HWHM, and $q$ is a phenomenological dimensionless shape parameter. The HWHM obtained is $\Gamma_d$ = 25.8$\pm$6 $\mu$eV which corresponds to $T_K$ = 300$\pm$70 mK.
The other parameters obtained from the fit are A = -5.5$\pm$0.7  $\mu$S, B = 0.11$\pm$0.06 $\mu$S, q = 0.10$\pm$0.05.

The ambiguity in the extracted Kondo temperature has been observed earlier on N-QD-N devices too ~\cite{Kretinin2012,VanderWiel2000}~. Kretinin et al. ~\cite{Kretinin2012}~ improved on the earlier NRG formula by suggesting that analysing the zero energy state at the lowest temperature as a function of a normalised energy scale $\nu\equiv eV_{SD}/(k_BT_K )$ is a more reliable method to extract the correct Kondo temperature of the system. It was based on the argument that temperature dependence data might have additional features arising from non-Kondo origin incorporated in the zero-energy state as a result of which the extracted $T_K$ would not define the Kondo state alone. The argument might hold for the anomaly in the value of $s$ parameter we obtained using the empirical formula fit. 

Using the equation derived by Kretinin et al.:
\begin{equation}
\label{eq_Kretinin}
\centering
    G(T=0,\nu)= \frac{G_{max}} {\bigg[1+\bigg(2^{\nicefrac{1}{s}}-1\bigg)\frac{\nu^2}{\pi} \bigg]^{s}}
\end{equation} 
and assuming our base temperature is close to the condition T = 0, we plot our 28 mK data (Figure 3(d)) along with two curves generated using Equation~\eqref{eq_Kretinin}. We notice that the curve generated by the $T_K$ and $s$ values obtained from the temperature dependence fit (blue dash dot) doesn't replicate the behaviour of the measured data. However, for the values $T_K$ = 270 mK (23.26 $\mu$eV) and $s$ = 0.22 (red dash dot), the curve closely follows the measured data.

\section{Discussion}
The results presented so far suggest that the zero-bias conductance peak we observe is indeed associated with a Kondo feature. Yet the larger gap $\Delta_1$ = 1.25 mV is far greater than $k_BT_K$, regardless of the method we choose to evaluate $T_K$.  
Although theoretical models do suggest that such a co-existence is possible, with a zero-bias feature retained even when $k_BT_K \ll \Delta$ ~\cite{Clerk2000,Cuevas2001,Tanaka2007,Domanski2008}~, most experiments carried out so far on N-QD-S and S-QD-S systems by various groups ~\cite{Buitelaar2002,Buizert2007,Yeyati2003,Sand-Jespersen2007,Graber2004,Deacon2010a,Deacon2010b,Kanai2010}~ do not observe Kondo features at this limit. 

Here we suggest that the co-existence between the Kondo effect and superconductivity is a consequence of the 2-gap nature of \nbse.
Since we observe the Kondo zero-energy state even at fields far from critical fields of bulk 2H-\nbse, $H_{c2}^{\perp} \approx $ 4 T, $H_{c2}^{\parallel} \approx$ 17 T~\cite{Xi2016}~, the normal electrons must be associated with a superconducting band that turns normal even at such low fields. 
As discussed previously, bulk \nbse has two superconducting gaps - one associated with niobium based bands around the $\Gamma$ and $K$ points and a second proximitized inner gap from the small selenium based bands around the $\Gamma$ point ~\cite{Dvir2018a,Kiss2007}~ whose signatures are observed at 1.25 mV and 0.6 mV respectively. 

To elucidate the role of the Se-derived band in mediating the Kondo effect, we examine the correlation between the appearance of a zero-energy state and the second band in out-of-plane magnetic field (Figure 4). The out-of-plane magnetic field is chosen since the inner gap exhibits stronger response to its onset~\cite{Dvir2018b}~. Figure 4(a) shows the zero-bias conductance as a function of $B_\perp$ as it is swept in the positive direction (indicated by the red arrow). A complete dataset tracing the zero-bias conductance $G_0$ from negative to positive fields is shown in Supplementary Figure 1. The conductance reaches the lowest value of 0.16 $\mu$S at $B_\perp$ = -0.2 mT, where the tunneling spectrum reaches a sub-gap conductance typical of a hard-gap \nbse junction ~\cite{Dvir2018a}~. At this limit, there is no indication of a zero-bias conductance peak.

As $B_\perp$ is increased, $G_0$ increases in discrete steps of $\approx$ 1 $\mu$S, observed at 1.0 mT, 2.2 mT and 3.1 mT. As $B_\perp$ is further increased, the $G_0$ increase becomes continuous. Tunneling spectra are measured for fields that mark a discrete increase in $G_0$ (Figure 4(b)) and also at two higher fields. The second derivative of tunneling current as a function of magnetic field is plotted in Figure 4(c). The data in Figure 4(b)-(c) are offset along the $y$-axis for clarity. We observe that the first signature of a distinct zero-bias peak appears at a field of 3.1 mT and persists at higher fields. Simultaneously, the feature at 0.6 mV corresponding to the Se derived superconducting bands loses prominence until it is difficult to resolve from its background.

\begin{figure}[ht]
    \centering
    \includegraphics[width = 400 pt]{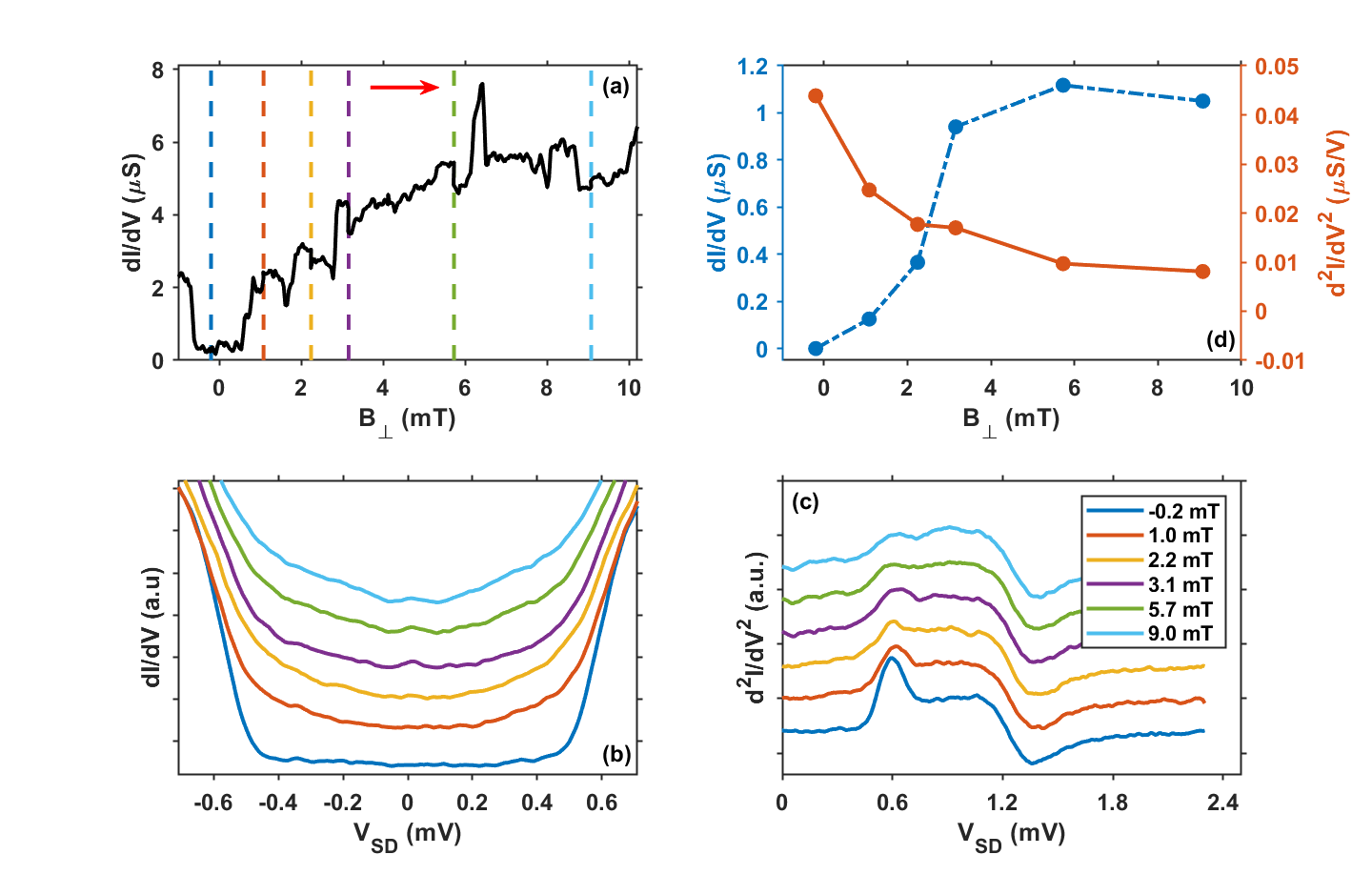}
    \caption{(a) Differential conductance at zero-bias ($G_0$) as a function of $B_\perp$ at mT resolution. The red arrow indicates the sweep direction of the $B_\perp$; (b) sub-gap tunneling spectra at various field values indicated by the coloured dotted lines in (a); (c) second derivative of the tunneling current at field values indicated by dotted lines in (a); (d) zero-bias conductance peak height (left $y$-axis) and height of the feature at 0.6 mV in second derivative of tunneling current (right $y$-axis) as a function of $B_\perp$}
    \label{fig:fig4}
\end{figure}

Figure 4(d) shows the evolution of the zero-bias conductance peak height (left $y$-axis) and the height of the feature in the second derivative of tunneling current (right $y$-axis) corresponding to the inner Se superconducting band at 0.6 mV. An inverse correlation is evident from the data: The peak at 0.6 mV is the strongest at -0.2 mT where the conductance at zero-bias reaches the lowest value. With reduction in height of the feature in magnetic field, the zero-bias conductance feature grows in strength. The reduced height of the zero-bias conductance peak at higher fields can be attributed to the higher background conductance with magnetic field. Supplementary Figure 2 shows the individual sub-gap tunneling spectra shown in Figure 4(b) with their background conductance. The discrete jumps ($\approx 1\mu$S) in the zero-bias conductance with magnetic field can be associated with the discrete entry of vortices in the tunnel junction area of \nbse ~\cite{Dvir2018b}~. This correlation thus supports the interpretation that the Se band turning  normal could assist in the spin-flip co-tunneling event, as depicted in the schematic Figure 5(a). 
Interestingly, we have shown in a previous work that in the weak coupling limit, a defect-bound QD favors coupling to the Se-band over coupling to the Nb-bands~\cite{Devidas2021}~.

\begin{figure}[ht]
    \centering
    \includegraphics[width = 400 pt]{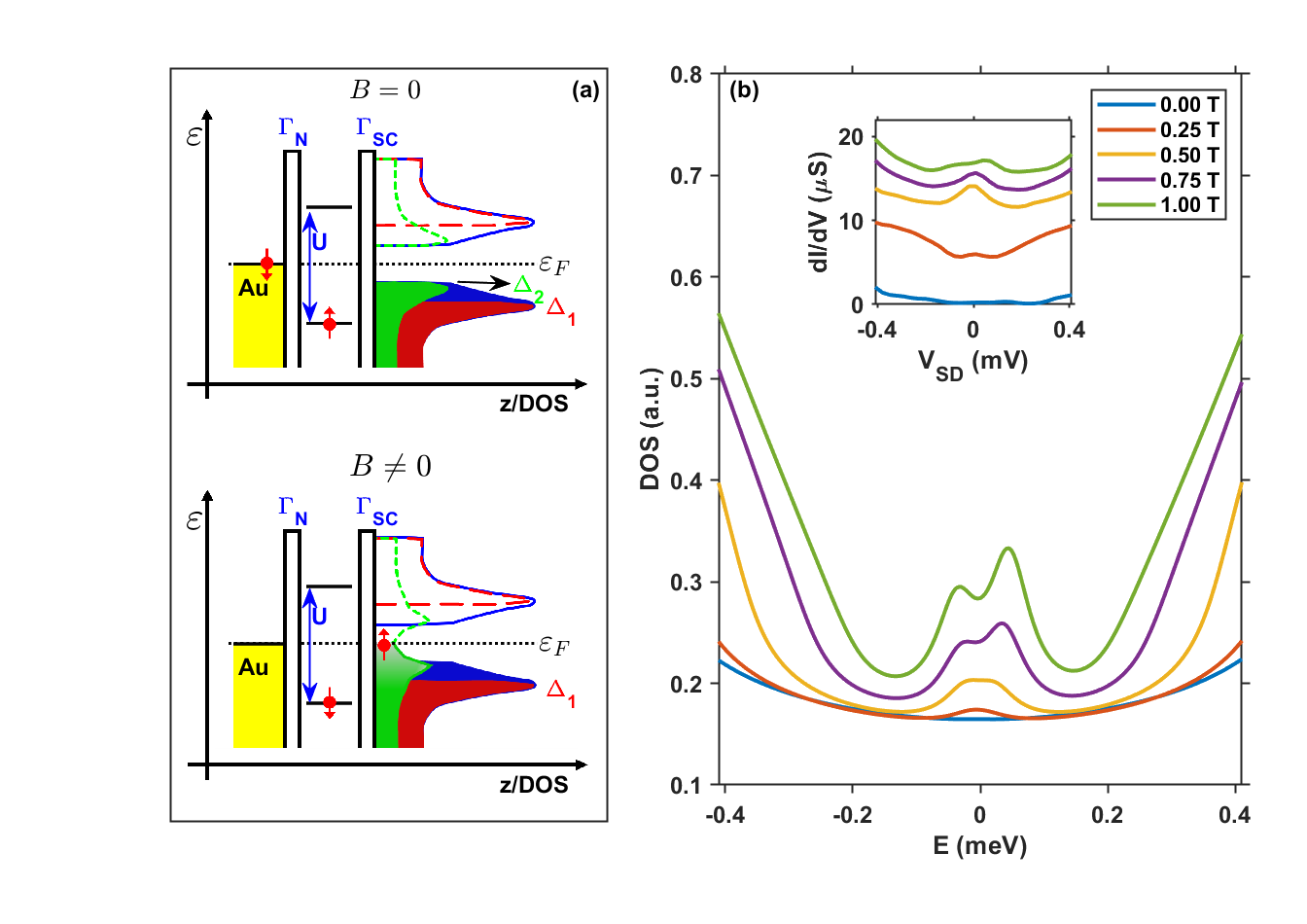}
    \caption{(a) Schematic of spin-flip co-tunneling event – (above) prohibited at zero applied magnetic field due to absence of states within the superconducting gap, (below) allowed via filled second band in \nbse in an applied magnetic field; (b) Sub-gap density of states of \nbse as a function of $B_{||}$, obtained via a large N expansion with a broadening $\delta$ = 0.04 meV. The inset shows the experimental differential tunneling conductance data for the same values of $B_{||}$.}
    \label{fig:fig5}
\end{figure}

To better understand the origin of the observed zero-bias conductance peak we 
study a Kondo model in which we 
treat the weakly coupled QD
as an impurity, with magnetic moment $\ssb$, placed at position $\bR$.
The Hamiltonian for the electrons in \nbse can
be written as $H=H_0+H_J$, where $H_0$ is the Hamiltonian
when no QD is present, and $H_J$ describes the coupling of the QD to the electrons in \nbse.
In the Kondo limit
$H_J= J \sum_{\sigma\sigma'} {c}^\dagger_{\bR\sigma}{\bm\tau}_{\sigma\sigma'}{c}_{\bR\sigma'}\cdot\ssb$, where
$J>0$ is the antiferromagnetic coupling between the QD and NbSe$_2$ electronic states, $\tau_i$ are the Pauli 
matrices in spin space,
${c}^\dagger_{\bR\sigma}$ and ${c}_{\bR\sigma}$ are the creation and annihilation operators, respectively, for an electron at position $\bR$ with spin $\sigma$.
To treat the interaction between $\ssb$ and the electrons we use the large-N expansion~\cite{Read1983,Bickers1987}~.
The large-N expansion has been shown to give accurate results, in agreement with other methods such
as the numerical renormalization group~\cite{bulla2008a,gonzalez-buxton1998}~, 
when the impurity is isolated and the Kondo problem has only one channel.
The electrons' density of states $\rho(\varepsilon)$ is the key property that determines the features of the Kondo effect.
Here, instead of assuming a constant $\rho(\varepsilon)$, or a prescribed form~\cite{Rossi2006,Mastrogiuseppe2014,Principi2015}~,
we use the density of states extracted for different values of the magnetic field
from spectroscopic measurements on few-layer superconducting \nbse~\cite{Dvir2018a}~.
One important aspect of using the experimentally obtained $\rho(\varepsilon)$ is that it allows us
to correctly capture the effect of $B$ on it, and therefore the unusual evolution
of the Kondo peak with $B$.
%
In the large-N expansion $\ssb$ is expressed in terms 
of auxiliary creation (annihilation) fermionic operators
${f}^\dagger_\sigma$ (${f}_\sigma$)
satisfying the constraint
$n_f=\sum_\sigma {f}^\dagger_\sigma {f}_{\sigma}=1$,
with  $\sigma = 1,\ldots, N$, so that
$H_J = J \sum_{\kk,\kk',\sigma,\sigma'} {c}^\dagger_{\kk\sigma}{c}_{\kk'\sigma'}{f}^\dagger_{\sigma'}{f}_\sigma$, with 
${c}^\dagger_{\kk\sigma}$ (${c}_{\kk\sigma}$) the creation (annihilation) operators for 
an electron with momentum $\kk$ and spin $\sigma$.
In the remainder we assume $|\ssb|=1/2$ and therefore set $N=2$.
We decouple the quartic interaction term $H_J$ via the mean-field 
$m\sim \sum_{\kk,\sigma} \langle  {\hat f}^\dagger_{\sigma} {\hat c}_{\kk\sigma} \rangle$.
 
The constrain $|\ssb|=1/2$ implies $n_f=1$ and this is enforced via a Lagrange multiplier, $\mu_f$, 
which plays the role of the chemical potential of the $f$-electrons.
By minimizing the effective action corresponding to $H=H_0+H_J$, within the saddle-point approximation,
we obtain~\cite{Bickers1987}~
\begin{align}
 &\sum_{\sigma=\pm 1}\int_{-D}^{D}d\varepsilon n_F(\varepsilon)
 \frac{\rho(\varepsilon)(\varepsilon-\mu_f+h\sigma)}{(\varepsilon-\mu_f+h\sigma)^2+[\pi\rho(\varepsilon)m^2]^2}+\frac{2}{J}=0; 
 \label{eq-m} \\
 &\sum_{\sigma=\pm 1}\int_{-D}^{D}d\varepsilon n_F(\varepsilon)
 \frac{\rho(\varepsilon)m^2}{(\varepsilon-\mu_f + h\sigma)^2 + [\pi\rho(\varepsilon)m^2]^2} -  1 = 0
 \label{eq-muf}
\end{align}
where $n_F(\varepsilon)$ is the Fermi function, 
$D$ is the electrons' bandwidth, and 
$h\approx (1/2)g_f\mu_B B$,  describes the Zeeman effect, with $g_f$ the effective g-factor for the QD
extracted from the experimentally observed Zeeman splitting.
Equations~\ceq{eq-m},~\ceq{eq-muf} can be inverted to find
$m^2$ and $\mu_f$ for a given value of $J$ and $B$.

From Eqs.~\ceq{eq-m},~\ceq{eq-muf} we obtain the values of $m$ and $\mu_f$ in the limit of $T\to 0$ (we set $T=6\times10^{-4}$~K).
For $\varepsilon\gg\Delta_1$, $\rho(\varepsilon)$, to very good approximation, can be taken to be constant.
Let $\rho_0$ be the value of $\rho(\varepsilon)$ for $\varepsilon\gg\Delta_1$.
The value of $\rho_0$ that enters Eqs.~\ceq{eq-m},~\ceq{eq-muf} depends on the effective spatial extension of the magnetic
impurity, i.e., in our case, of the defect-bound QD.
$\rho_0$ and $J$ are very difficult to estimate, both theoretically and experimentally.
On the other hand, $m$ and $\mu_f$, depend mostly on the product $\rho_0 J$,
and not on the separate values of $\rho_0$ and $J$. We studied the dependence of 
the Kondo peak on the product of $\rho_0 J$ and found the general result that the peak grows with magnetic field for
$\rho_0 J\lesssim 0.2$ and decreases for $\rho_0 J\gtrsim 0.2$.
We find that the scaling of the Kondo peak with $B$ observed experimentally is best approximated
when $\rho_0 J\approx 0.14$. In the remainder we therefore set $\rho_0 J = 0.14$.

We obtain the electrons' Green's function at position $\bR$ and energy $\omega$, $G(\bR,\bR,\omega)$, renormalized by the coupling to the effective magnetic impurity, via the equations
\begin{align}
 & G(\bR,\bR,\omega) = G_0(\bR,\bR,\omega) + m^2 G_0(\bR,\bR,\omega) F(\omega) G_0(\bR,\bR,\omega),
 \label{eq-G} \\
 & F(\omega)=[(\omega+i\delta-\mu_f-m^2 G_0(\bR,\bR,\omega))\tau_0-(1/2)g\mu_B B\tau_z]^{-1},
 \label{eq-F}
\end{align}
where
$G_0(\bR,\bR,\omega)=-i(\pi/2)\rho(\omega)\tau_0$,
is the electrons' bare local Green's function,
$F(\omega)$
is the Green's function for the effective fermionic degrees of freedom $f$, and $\delta$ is 
a small broadening that we introduce to take into account the effect on the spectra of thermal fluctuations and non-magnetic disorder.
The local DOS at the position of the QD, at energy $\omega$, $\rho_{\rm tot}(\omega)$ is then obtained as 
$\rho_{\rm tot}(\omega)=-{\Im}[{\rm Tr}(G)]/\pi$. 

Figure~\ref{fig:fig5}~(b) shows  $\rho_{\rm tot}(\omega)$ for different values of $B$ obtained via the large-N expansion assuming 
$\delta=0.04 ~\text{meV}$. We see that for $B=0$, given the vanishing of the DOS as $\varepsilon\to 0$ for superconducting \nbse, no peak for $\varepsilon\approx 0$ is present in $\rho_{\rm tot}(\omega)$, indicating the absence of Kondo screening.
NRG results~\cite{Ingersent1996}~ show that when $N(\varepsilon)\sim |\varepsilon|^a$ with a $>$ 1/2, and perfect particle-hole symmetry is present, no Kondo effect can be realized.
In superconducting \nbse the presence of a magnetic field $B$ induces a finite density of states at low energies.
This makes possible the establishment of a Kondo cloud, and, when $B$ is not too large,
dominates over the suppression of $T_K$ due to the polarization of $\ssb$ induced by the magnetic field.
For $B\gtrsim 0.25$~T a peak develops: the softening, at low energies, of the superconducting
gap allows the establishment of Kondo screening. At larger $B$ the Kondo peak splits, as expected~\cite{Hewson1993}~.

\section{Summary}
Our work shows that defect-bound QDs can be used as platforms to probe the interplay between the energy scales associated with the Kondo effect and with superconductivity. This opens a number of interesting future possibilities. First, as discussed above, \nbse is a 2-gap superconductor - offering a richer phase space involving both order parameters. Interestingly, \nbse is expected to develop a triplet order parameter at high $B_\parallel$~\cite{mockli2020}~. This could give rise to a field-driven transition in the type of Kondo effect. Finally, it is interesting to consider coupling QDs to unconventional SCs such as FeTe$_{0.55}$Se$_{0.45}$~\cite{Zalic2019}~, which can be produced using exfoliation, or even 
to the superconducting states appearing in twisted graphene systems~\cite{Cao2018a}~.  

\section{Methods}
The tunneling measurements are performed using standard low frequency lockin techniques in a BlueFors dilution cryostat with a base temperature of 20 mK. An AC excitation of 30$\mu V$ is applied across the device between the tunneling electrodes and the ohmic drain electrode using a Zurich instruments MFLI digital lock-in amplifier (LIA). The superconductor is tuned in and out of the superconducting gap by applying a DC bias voltage ($V_{SD}$), also obtained from the LIA. The tunneling current reaching the drain is fed into a FEMTO current amplifier. The output of the current amplifier is then fed into a Keithley 2000 digital multimeter to measure $I_{DC}$ and into the LIA input to measure the differential conductance $dI/dV$ as a function of $V_{SD}$. The magnetic field measurements are performed with the 2 axis 9T-3T (Z-Y) superconducting magnet attached to the dilution. The magnet is powered by the commercial power supply from American Magnetic Inc. (AMI), with a resolution of 1 mT (supplied). For resolutions smaller than that, the magnet is powered by a Keithley 2400 SMU in the current source mode. The applied field is evaluated from the current to field ratio provided by AMI.

\section{Acknowledgements}
Devices for this study were fabricated at the Center for Nanoscience and Nanotechnology, The Hebrew University.
Funding for this work was provided by a European Research Council Starting Grant (No. 637298, TUNNEL), Israel Science Foundation grant 861/19, and BSF grant 2016320.
E.R.  thanks the Aspen Center for Physics, which is supported by National Science Foundation grant PHY-1607611, where part of this work was conducted, and Kevin Ingersent for very helpful discussions.

\bibliography{references} 
\beginsupportinginfo
\newpage

\section*{Supplementary Information}
\subsection*{Supplementary Figure 1}
\begin{figure}[H]
    \centering
    \includegraphics[width=325pt]{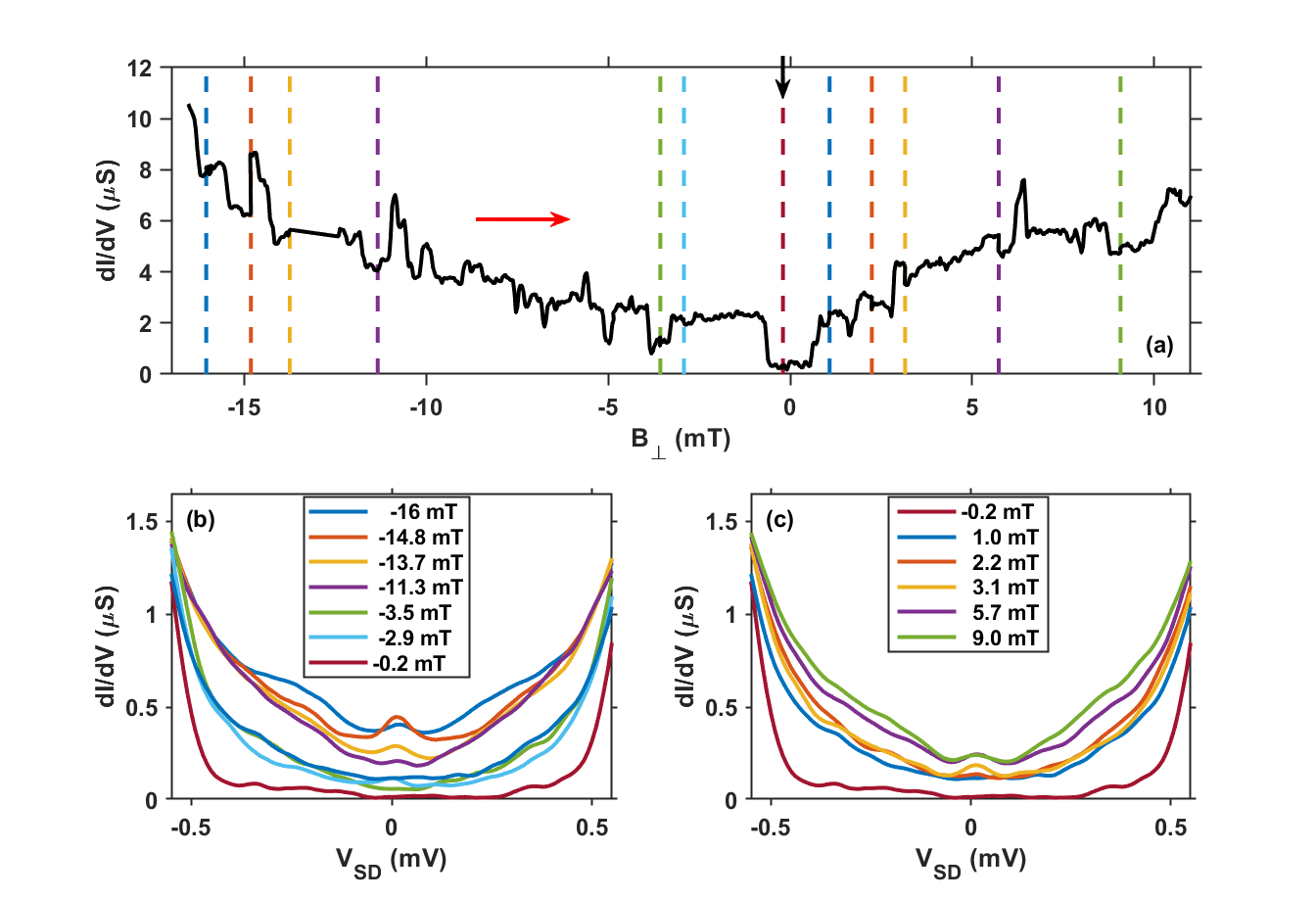}
    \caption{(a) Differential conductance at zero bias ($G_0$) as a function of $B_\perp$ at mT resolution over an extended range. Red arrow indicates the scan direction. $G_0$ is the lowest at -0.2 mT, indicated by a black arrow; Sub-gap differential conductance spectra dI/dV for magnetic field values (b) lower than -0.2 mT and (c) higher than -0.2 mT are shown. The colored dash-lines indicate the magnetic field at which the spectrum was measured.}
\end{figure}

\newpage
\subsection*{Supplementary Figure 2}

\begin{figure}[H]
    \centering
    \includegraphics[width=380pt]{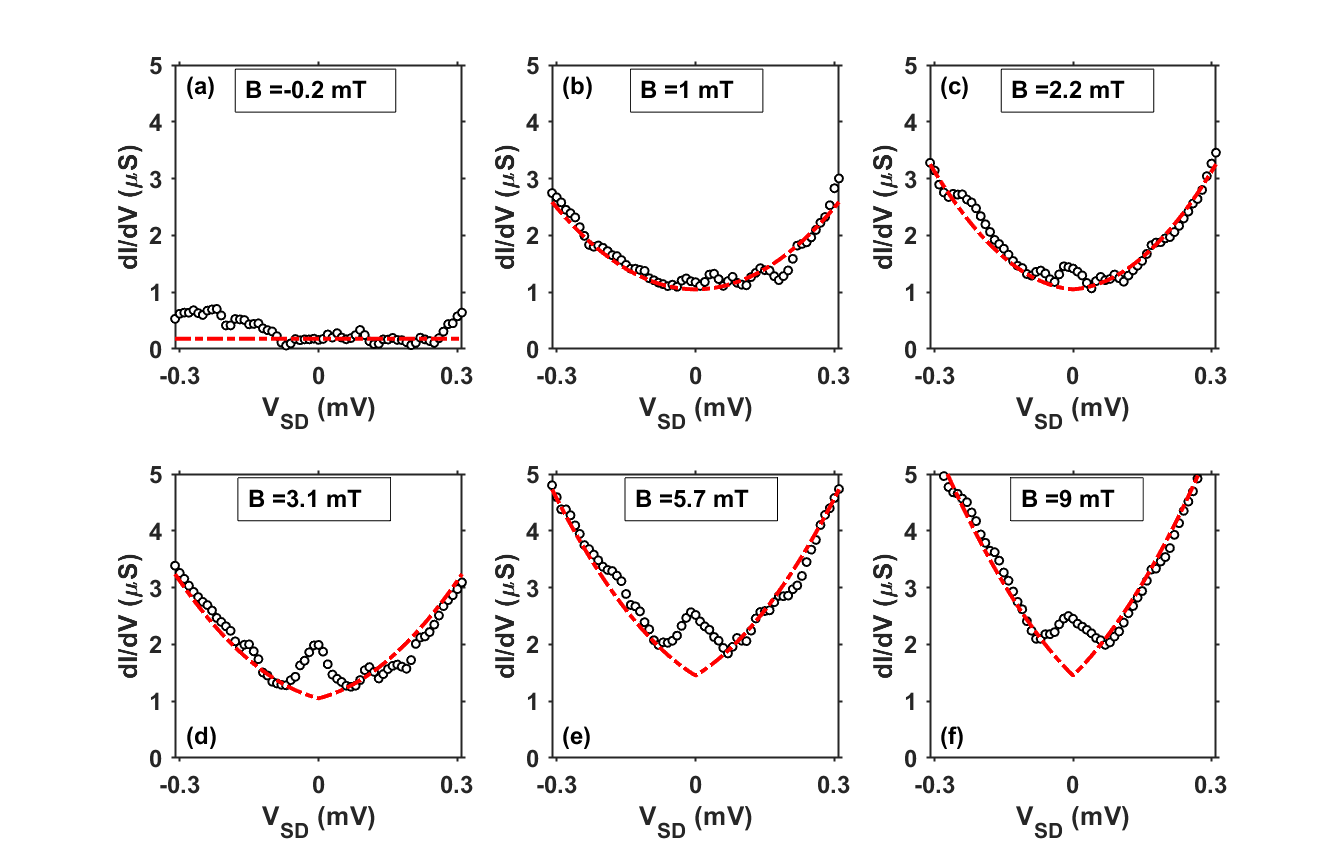}
    \caption{(a)-(f) Sub-gap differential conductance (dI/dV) spectra of QD-\nbse junction along with the respective background conductance.}
\end{figure}

The individual sub-gap differential conductance spectra of the QD-\nbse junction, for the magnetic fields shown in Figure 4 in main text, in the range $-0.3 \text{ mV} \leq \Delta \leq 0.3 \text{ mV}$ are shown in Figure S2. The red dot-dash lines are the background conductance estimated using the function $a_0+\alpha \abs{x}+\beta x^2$; where $a_0$ is the conductance at $V_{SD} = 0 \text{ V}$, $\alpha$ and $\beta$ are coefficients. The values for the $a_0$, $\alpha$ and $\beta$ are tabulated in Table \ref{tab:Background conductance}.

\vspace{1pt}
\begin{table}
    \centering
    \begin{tabular}{|c|c|c|c|}
    \hline
    \thead{Magnetic Field} & $a_0$ & $\alpha$ & $\beta$ \\
    (mT) & ($\mu S$) &  & \\[1pt]
    \hline
    \hline 
    \multirow{1}{*} -0.2 & 0.16 & 0 & 0 \\
    \hline
    \multirow{1}{*} 1.0 & 1.05 & 0 & 16 \\
    \hline
    \multirow{1}{*} 2.2 & 1.05 & 0.6 & 21 \\
    \hline
    \multirow{1}{*} 3.1 & 1.05 & 2.1 & 16 \\
    \hline
    \multirow{1}{*} 5.7 & 1.45 & 5 & 18 \\
    \hline
    \multirow{1}{*} 9.0 & 1.45 & 8 & 18.5 \\
    \hline
    \end{tabular}
    \caption{Coefficients used to obtain the background conductance for $dI/dV$ spectra in Figure S2(a)-(f)}
    \label{tab:Background conductance}
\end{table}

\end{document}